\documentclass[12pt]{article}

\usepackage[dvips]{graphicx}
\usepackage{a4wide}

\newcommand{\nn}{\nonumber}
\newcommand{\be}{\begin{equation}}
\newcommand{\ee}{\end{equation}}
\newcommand{\ba}{\begin{eqnarray}}
\newcommand{\ea}{\end{eqnarray}}
\newcommand{\req}[1]{(\ref{#1})}
\newcommand{\ci}[1]{\cite{#1}}
\newcommand{\da}{{distribution amplitude}}

\def\gev{~{\rm GeV}}

\def\ale{\alpha_{\rm elm}}

\def\muF{\mu^2_F}
\def\muR{\mu^2_R}

\def\ub{\bar{u}}
\def\db{\bar{d}}
\def\sb{\bar{s}}
\def\sh{{s}}
\def\uh{{u}}
\def\th{{t}}
\def\pb{\bar{p}}
\def\qb{\bar{q}}
\def\={\,=\,}
\newcommand{\sla}{\hspace*{-0.20cm}/}

\begin{document} 
\thispagestyle{empty}
\begin{flushright}
WU B 05-04 \\
hep-ph/0505258\\
May 2005\\[20mm]
\end{flushright}

\begin{center}
{\Large\bf The process $p\pb\to\gamma\,\pi^0$ within the handbag approch} \\
\vskip 15mm

P.\ Kroll$^a$ and 
A.\ Sch\"afer$^b$\\[1em]
a) {\small {\it Fachbereich Physik, Universit\"at Wuppertal, D-42097 Wuppertal,
Germany}}\\
b) {\small {\it Institut f\"ur Theoretische Physik, Universit\"at Regensburg,
D-93040 Regensburg}} 
\vskip 5mm

\begin{abstract}
We analyse the exclusive channel $p\pb\to\gamma\,\pi^0$, assuming handbag 
dominance. The soft parts are parametrized in terms of CGLN
amplitudes for the $q\qb\to\gamma \pi^0$ transition and form factors for the
$p\pb\to q\qb$ ones, the latter represent moments of Generalized 
Distribution Amplitudes. We present a combined fit to Fermilab data 
from E760  taking simultaneously into account information from other 
exclusive reactions, especially from $p\pb\to \gamma\gamma$
data. Overall a nicely consistent picture emerges, such that one can
hope, that our  theoretical analysis will be reliable also for the
kinematics of GSI/FAIR, which hopefully will provide much more precise
and complete data. Consequently, data from this facility should
improve our knowledge both on the proton-antiproton distribution 
amplitudes and the pion production mechanism. 
\end{abstract}
\end{center}

\section{ Introduction}
The reliable theoretical treatment of hard exclusive processes has been a 
challenge for QCD for many years. With the advent of the 
generalized parton distribution formalism 
\cite{GPDs} a large class of such processes, all involving some hard 
scale $Q^2$, can now be treated on a firm, perturbative  QCD basis,
absorbing all non-perturbative soft physics in suitable generalized parton
distributions. This success is made possible by the dominance
of the handbag diagrams in leading twist, as demonstrated in the  
factorization proofs.\\
There are other hard exclusive processes for which these rigorous
proofs do not apply.
Still, however, there are good arguments for the dominance of the 
handbag contribution in certain regions of phase space 
also for many of these reactions. Examples of such processes are
two-photon annihilations into pair of hadrons for large but not
asymptotically large Mandelstam variables, $s$, $-t$, $-u$, for
applications see \ci{DKV2,DKV3,freund:02}.\\
The FAIR project at GSI with the HESR anti-proton program \cite{PAX}
will offer ideal possibilities to study exclusive channels
in $p\bar p$ annihilation, e.g. $p\bar p \rightarrow
\gamma \gamma$, the time reversed of the widely studied 
process  $\gamma \gamma\rightarrow p\bar p$ \cite{CLEOp,BELLE}.
Another very interesting channel is $p\bar p \rightarrow \pi^0 \gamma$ 
because rates will be much higher and the amplitude is related by 
crossing to meson photoproduction $p\gamma \rightarrow \pi^0 p$. 
The latter process has been recently investigated within the handbag
approach in Refs.~\ci{huang:00,huang:04}. It turned out that a 
leading-twist calculation of the partonic subprocess, which is
$\pi^0$ photoproduction off quarks, is insufficent. In leading-twist 
accuracy, i.e.\ considering only the one-gluon exchange mechanism for 
the generation of the meson, the resulting $\gamma p\to \pi^0 p$ cross 
section is far too small. This parallels observations made in
leading-twist calculations of the pion form factor \ci{pion}. The
results are typically a factor 3 to 4 below the admittedly poor data
available at present. In fact the lowest-order Feynman graphs
contributing to $\gamma p\to \pi^0 p$ within the handbag approach 
are the same as those occuring in the calculation of the pion form
factor. Therefore, one has to conclude that for the existing data one 
is still far from the asymptotic region in which one-gluon exchange 
dominates and it is necessary to use a more general mechanism for the 
generation of the meson. Many alternatives and/or corrections to the 
leading-twist  meson generation have been discussed in the literature 
reaching from higher-twist or power corrections to resummation of
perturbative corrections. Still the description of meson generation is 
a matter of controversy. In order to remedy the situation a treatment 
of the subprocess is called for that does not postulate the dominance  
of any specific meson generation mechanism. Huang et al. \ci{huang:04} 
have proposed such a method. They utilized the covariant decomposition 
of the $\gamma q\to \pi^0 q$ amplitudes proposed by Chew, Goldberger, 
Low and Nambu (CGLN) \ci{CGLN}. This decomposition separates the
kinematic and helicity dependences of the subprocess from the dynamics 
of the meson generation which is embodied in the CGLN invariant
functions. Exploiting properties of the invariant functions various 
dynamical mechanisms can be identified and consequences discussed. 
Comparison with experimental data on the ratio of $\pi^+$ and $\pi^-$ 
photoproduction cross sections \ci{JLab1} provides evidence for the 
dominance of one invariant function out of the set of four.\\ 
In this work we are going to investigate the corresponding time-like process
$p\pb\to\gamma\,\pi^0$. As for its space-like partner we will again make
use of the CGLN decomposition as at least for the fixed-target option
of FAIR ($s<30$ GeV$^2$) but possibly also for the collider option 
($s<210$ GeV$^2$) the one-gluon exchange mechanism for meson
generation will not dominate. Comparison with existing data from
Fermilab \ci{E760:97} at 8.5 GeV$^2$ $\leq s \leq $ 13.6 GeV$^2$
allows for a critical examination of the handbag charateristics. We
will show that, in parallel to the space-like region, there are
indications for the dominance of one of the invariant functions. In
fact, this function is the $s\leftrightarrow t$ crossed one of the
seemingly leading one in the space-like region. Morever, this function
is the one that is fed by the leading-twist mechanism although not
sufficiently strongly. Exploiting the crossing properties of the
CGLN invariant functions, we are in the position to relate the space-
and time-like processes quantitatively. Let us note that for the
complementary kinematic regime, namely production of a virtual
photon and a pion into the forward direction,  a different
factorization scheme was proposed by Pire and Szymanowski \ci{pire}.

\section{The handbag amplitude for $p\pb\to \gamma\,\pi^0$}
The treatment of proton-antiproton annihilation into photon and meson  
parallels that of annihilation into two photons respectively its
time-reversed process, two-photon annihilation into $p\pb$,
investigated in \ci{DKV3}. Therefore, we can take over many of the results 
derived in that publication and for clarity we will also use a notation 
as close as possible to that employed in \ci{DKV3}.\\
Obviously our crucial starting point is the assumption of  handbag 
factorization of the amplitude for the kinematical region 
$s, -t, -u \gg \Lambda^2$ where 
$\Lambda$ is a typical hadronic scale of the order of 1 \gev. In this
factorization scheme for which validity arguments have been given in
Ref.~\ci{DKV3}, the process amplitudes appear as a product of a hard
subprocess, $q^a\qb^a \to \gamma\,\pi^0$ and a soft $p\pb\to q\qb$ 
transition matrix element which is parametrized by $p\pb$ distribution 
amplitudes $\Phi^a_i$, $i=A,P,V,S$ introduced in \ci{DKV3} and where 
$a$ is the  flavour of the quark anti-quark pair  emitted from the
$p\pb$ pair. The distribution amplitudes  $\Phi^a_i$ are time-like 
versions of general parton distributions for the proton.\\
We are interested in large-angle scattering processes. 
For very high energies the handbag diagram will not dominate, but rather 
processes like that in Fig. 1a \ci{brodsky}.
However, this is not a consequence of power counting, but rather of Sudakov 
suppression, which can only be expected to be effective at really large 
energies. (A nearly real parton entering or leaving a hard scattering 
process emits gluon-bremstrahlung. Excluding this part of the cross section 
by requireing  exclusivity therefore leads to suppression.) 
See \cite{DFJK4} for details. The asymptotic contribution a l\'a
Brodsky and Lepage \ci{brodsky} to
our process has not yet been worked out. Experience with other
processes like $\gamma\gamma\to p\pb$ \ci{farrar} or $\gamma
  p\to\gamma p$ \ci{dixon} let us, however,
expect that this contribution to the cross section is likely 
to lie way below the experimental points for $s$
of the order of $10\,\gev^2$.\\
\begin{figure}[t]
\begin{center}
\includegraphics[width=12 cm]{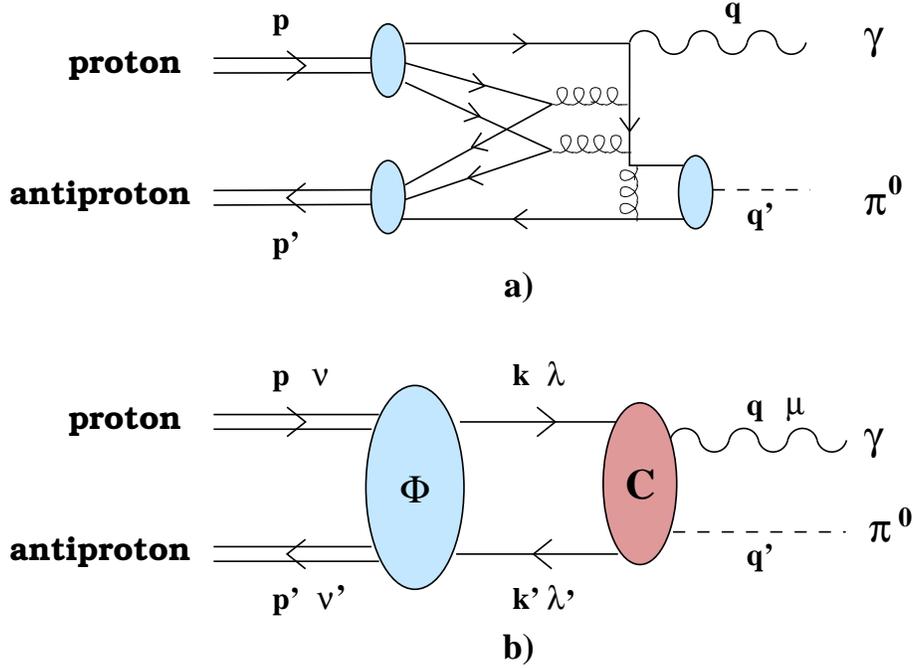}
\end{center}
\caption{a) One of the leading-twist contributions at asymptoticaly
large $s$. The two blops on the left-hand side represent ordinary
proton distribution amplitudes while the one on the right-hand side
is the pion distribution amplitude. b) The handbag contribution at
large but non-asymptotic $s$. The blob at the left-hand side
represents the $p\pb$ distribution amplitude, the one on the
right-hand side the general dynamics for $q\qb\to \gamma\pi^0$
parametrized in terms of the CGLN invariant functions. The momenta
and helicities of the various particles are indicated. }
\end{figure}

As we will follow closely the discussion of 
two-photon annihilation into $p\pb$, presented in \ci{DKV3}, we 
refrain from presenting all details of the theoretical analysis, and
rather restricting ourselves to a sketch of the main points. We work
in a symmetric frame where the momenta of the proton and the antiproton 
are defined in light-cone coordinates as
\be
p \= \sqrt{\frac{s}{8}}\, [1,1,\sqrt{2}\,\beta\, {\bf e}_\perp]\,, \qquad
p^\prime \= \sqrt{\frac{s}{8}}\, [1,1,-\sqrt{2}\,\beta\, {\bf e}_\perp]\,, 
\ee
where $\beta\=\sqrt{1-4m^2/s}$ and $m$ is the mass of the proton. 
In the following we will work in the massless limit. The transverse
direction is characterized by the two-dimensional vector ${\bf e}_\perp$
for which we choose $(1,0)$. The important feature of a symmetric frame 
in the time-like region is that the skewness is
\be
\zeta\=p^+/(p^+ + p^{\prime +}) \= 1/2\,.
\ee
The momenta of the photon and the pion read
\be
q \= \sqrt{\frac{s}{8}}\, [1+\sin{\theta},1-\sin{\theta},
             \sqrt{2}\,\cos{\theta}\, {\bf e}_\perp]\,, \quad
q^\prime \= \sqrt{\frac{s}{8}}\, [1+\sin{\theta},1-\sin{\theta},
             -\sqrt{2}\,\cos{\theta}\, {\bf e}_\perp]\,.
\ee
in our frame of reference which is also a c.m. frame.
The scattering angle is denoted by $\theta$ and the trigonometric
functions of it are related to the Mandelstam variables by
\be
\sin{\theta} \= \frac{2\sqrt{tu}}{s}\,, \qquad 
\cos{\theta} \= \frac{t-u}{s}\,.
\label{sin-cos}
\ee 
The starting point for the derivation of the handbag amplitude is the
expression
\be
{\cal M} \= \sum_{a} e\, e_a  \int d^4k \int
  \frac{d^4x}{(2\pi)^4} {\rm e}^{-ikx}\, \langle\, 0\, |\,
                    T \bar{\Psi}^a_\alpha(0)\, \Psi^a_{\beta}(x)\,
                      |\,p(p)\, \pb(p^\prime)\,\rangle\, 
                    {\cal H}_{\alpha\beta}^a\,,
\label{start}
\ee
where we omitted helicity labels here for convenience.
The hard scattering kernel, describing the subprocess $q^a(k)\, \qb^a(k^\prime)\to
\gamma\, \pi^0$, is denoted by ${\cal H}^a$. The sum runs over the quark
flavours $a\=u,d,s$; $e_a$ is the corresponding charge in units of
the positron charge $e$. As discussed in detail in \ci{DKV2,DKV3,DFJK1},
the $p\pb\to q \qb$ transition at large $s$ can only be soft if the
outgoing quark and antiquark have small virtualities and each carries
approximatively the momentum of the baryon or antibaryon; the
deviations of the parton momenta from the hadronic ones in size and
direction are of order $\Lambda^2/s$. It can be further shown that
the dominant Dirac structure of the soft transition matrix element in
\req{start} involves the good components of the quark fields in
the parlance of light-cone quantization. The hard scattering
amplitudes can approximately be calculated with on-shell parton
momenta: $H^a=\bar{u}^a{\cal H}^a v^a$. This guarantees gauge invariance. 
Within our calculational scheme we also use the approximation that the
proton (antiproton) dominantly emits a valence quark (antiquark).
Contributions from the emission of fast sea quarks are expected
to be small.\\
Putting all this together we obtain, in full analogy to the case of
$p\pb\to\gamma\gamma$ \ci{DKV3}, the handbag amplitude for the process 
under consideration (for the sake of legibility explicit helicities
are labelled by their signs)
\ba
\label{handbag-amp}
{\cal M}_{\mu0,\nu\nu'}&=& \sum_a \frac{e}{2}\, e_a 
\left\{
         \Big[\,H^a_{\mu\, 0,+-} + H^a_{\mu\, 0, -+}\,\Big]\, \delta_{\nu -\nu'}\,
         F^{\,a\,*}_V(s) \right. \\ 
             & +& \left.\hspace*{-0.3cm}\Big[\,H^a_{\mu\, 0, +-}-
           H^a_{\mu\, 0,-+}\,\Big]\;  
                \Big[\,2\nu\, \delta_{\nu-\nu'}\, (F^{\,a\,*}_A(s)+F^{\,a\,*}_P(s)) 
                 - \frac{\sqrt{s}}{2m}\,\delta_{\nu\nu'}\,
                 F^{\,a\,*}_P(s) \Big]\right\}  + {\cal O}(\frac{\Lambda^2}{s})\,,
\nonumber
\ea
where we make use of the definition of the annihilation form factors
in terms of the $p\pb$ \da s $\Phi^a_i$ as introduced in \ci{DKV3}
\be
     F^{\,a}_i(s) \= \int_0^1 dz\; \Phi^a_i (z,\zeta,s,\muF)  \qquad {\rm for}
     \quad i\= V,A,P\,.
\label{moments}
\ee
In Ref.\ \ci{DKV3} the form factors are defined for the soft
transitions $q^a\qb^a \to p\bar{p}$ while here we consider the
time-reversed transitions. Therefore, the complex conjugated form
factors occur here. The fourth form factor or \da{}, the scalar one,
decouples in the symmetric frame ($\zeta=\frac{1}{2}$), see
\cite{DKV3}. Note that the relations \req{moments} hold for any
physical value of the skewness. They also hold for any value of the
factorization scale, $\muF$, of the \da s, since the vector and axial
vector currents have zero anomalous dimensions. Because of symmetry 
properties of the $p\pb$ \da s \ci{DKV3}, $F^{\,a}_A$ and $F^{\,a}_P$ 
project onto the C-even part of the $p\pb$ state while $F^{\,a}_V$ is 
related to the C-odd part. On the other hand, the $\gamma\,\pi^0$
state is C-odd. This apparent conflict is a consequence of neglecting 
configurations where the antiquark is emitted in the direction of the 
proton instead of the antiproton. According to the arguments presented 
in \ci{DKV3} this contribution is small.\\
We emphasize that quark-antiquark helicity non-flip does not occur in
\req{handbag-amp} although there is baryon helicity non-flip. Although
we deal with massless quarks, quark helicity non-flip could contribute
since we have not specified the mechanism that controls the
generation of the meson. While the leading-twist, one-gluon exchange
mechanism is pure helicity flip, twist-3 effects, for instance,
produce non-flip contributions. Such contributions which are accompanied by
helicity-non-flip generalized \da s and associated form factors
\ci{diehl:01},
have been studied for the corresponding space-like process in \ci{huang:04}. 
In this respect $p\pb$ annihilation into $\gamma\,\pi^0$ differs from 
annihilation into two photons \ci{DKV3}. The subprocess for the latter
reaction is pure helicity flip for massless quarks. In parallel to
pion photoproduction where quark helicity flip is neglected \ci{huang:04}, 
we assume that quark helicity non-flip contributions are negligible
in the time-like region. In principle, this conjecture can be
tested by measuring helicity correlations. As we argued above the
dominant contribution to our process comes from the emission of
valence (anti)quarks by the (anti)proton. We therefore have to take
into account only the two subprocesses $u\ub\to \gamma \,\pi^0$ and 
$d\db\to \gamma\,\pi^0$. By isospin symmetry the weights of these two 
subprocesses are ${\cal C}^u_{\pi^0}=1/\sqrt{2}$ and
${\cal C}^d_{\pi^0}=-1/\sqrt{2}$, see the discussion in \ci{huang:04}.
It is therefore convenient to pull out these weight factors from the
subprocess amplitudes, $H^a={\cal C}^a_{\pi^0}\, H$, and absorb them 
as well as the corresponding fractional charges into the form factors
by introducing, following previous work \ci{DKV3,huang:00,huang:04,DFJK1}, 
annihilation form factors specific to $p\pb$ annihilation into a
photon and a $\pi^0$ ($i=V,A,P$)
\be 
R_{\,i}^{\,\pi^0} = \frac{1}{\sqrt{2}} \Big(\, e_u\, F_i^{\,u} -
                                           \,e_d\, F_i^{\,d} \Big)\,, 
\label{pi-ff}
\ee
After this procedure the subprocess amplitudes are flavor independent
and we can drop the corresponding superscript.     

To proceed we make use of the CGLN covariant decomposition of the
subprocess amplitudes \ci{CGLN}:
\be
H_{\mu 0,\lambda\lambda'}\= \sum_{i=1}^4\, \overline{C}_i(s,t)
\bar{u}(k,\lambda)\, \overline{Q}_i\, v(k',\lambda')\,,
\ee
with the manifestly (elm.) gauge invariant covariants 
\ba
\overline{Q}_1 &=& -2\gamma_5\Big[ \bar{k}\cdot \epsilon^*\, q\cdot q'- \bar{k}\cdot
q\, q'\cdot \epsilon^*\Big]\,,\nn\\
\overline{Q}_2 &=& -2\gamma_5\Big[ \bar{k}\cdot q\, \epsilon\sla^* - \bar{k}\cdot
\epsilon^*\, q\sla\Big]\,,\nn\\
\overline{Q}_3 &=& -\phantom{2}\gamma_5\Big[q\cdot q'\, \epsilon\sla^* - q'\cdot
\epsilon^*\, q\sla\Big]\,,\nn\\
\overline{Q}_4 &=& -\phantom{2}\gamma_5\, \epsilon\sla^*\, q\sla\,.
\label{CGLN-cov}
\ea
encoding the helicity structure of the subprocess.
The vector $\epsilon$ denotes the polarisation of the photon and 
\be
\bar{k} \= \frac12\,(k-k')\,.
\ee
The invariant functions $\overline{C}_j$ depend on
the detailed, unknown  dynamics. Using the spinor definitions given in Ref.\
\ci{diehl:01}, Eq.\ (17), one can calculate the helicity amplitudes
in terms of these invariant functions. This reveals that $\overline{C}_1$ and
$\overline{C_4}$ only contribute to the parton non-flip amplitudes in the
massless limit which, as just discussed, will be neglected. The other
two invariant functions contribute to the helicity flip amplitudes
\ba
H_{+0,+-}(s,t) &=&
-\sqrt{\frac{s}{2}}~u \Big[ \overline{C}_2- \overline{C}_3 \Big]\,,
\nonumber \\[0.5em]
H_{+0,-+}(s,t) &=& \phantom{-} \sqrt{\frac{s}{2}}~t \Big[
  \overline{C}_2+ \overline{C}_3 \Big]\,.
\label{CGLN-amp}
\ea
The Mandelstam variables in the subprocess may differ
from the ones for the full process due to the proton mass, see for
instance \ci{DFHK}. Here, in this work we will ignore such possible
target mass corrections which are of order $m^2/s$. Inserting
\req{pi-ff} and \req{CGLN-amp} into \req{handbag-amp}, we obtain the
amplitudes
\ba
{\cal M}_{+ 0,\nu \nu'} &=& +\frac{e}{2}\, \sqrt{\frac{\sh}{2}}\, 
               \left\{\Big[\,+(\th-\uh)\, \overline{C}_2    
                         - \sh\,\overline{C}_3\,\Big]\,\delta_{\nu -\nu'}\,
                             R_V^{\,\pi^0\,*} \right. \nn\\
                  &+& \left. \Big[\, \sh\, \overline{C}_2 -(\th-\uh)\,
                         \overline{C}_3\,\Big]\, \Big[\,
              2\nu\,\delta_{\nu -\nu'}\, (R_A^{\,\pi^0\,*}+R_P^{\,\pi^0\,*}) -
                      \frac{\sqrt{s}}{2m}\,\delta_{\nu\nu'}\,
                      R_P^{\,\pi^0\,*}\,\Big]\right\}\,, 
\label{hb-amp}
\ea
which lead to the $p\pb\to\gamma\,\pi^0$ cross section
\ba
\frac{d\sigma}{dt}(p\pb\to\gamma\,\pi^0) &=&
       \frac{\ale}{32 
}\, \sh\,   \nn\\
    \hspace*{-1cm}  &\times& \left\{ |\overline{C}_2(\sh,\th)|^2\,
       \Big[\Big(\frac{\th-\uh}{\sh}\Big)^2\, |R_V^{\,\pi^0}(s)|^2 +
      (R^{\,\pi^0}_{\rm eff}(s))^2 \Big]\right.\nn\\  
       &+& \left.|\overline{C}_3(\sh,\th)|^2\,
       \Big[|R_V^{\,\pi^0}(s)|^2 +
          \Big(\frac{\th-\uh}{\sh}\Big)^2\,  (R^{\,\pi^0}_{\rm eff}(s))^2  
                   \Big ]\right.\nn\\ 
        &-& \left. 2 {\rm Re}\big[ \overline{C}_2(\sh,\th)
       \overline{C}_3^*(\sh,\th)\big]\, 
       \Big(\frac{\th-\uh}{\sh}\Big) \,\Big[|R_V^{\,\pi^0}(s)|^2 +
          (R^{\,\pi^0}_{\rm eff}(s))^2 \Big] \right\}\,.
\label{cross-section}
\ea
neglecting $\mathcal{O}(m^2/s)$ terms.
Since always a particular combination of the form factors
$R_A^{\,\pi^0}$ and $R_P^{\,\pi^0}$ appears in the cross section   
we introduce an effective form factor
\be
R^{\,\pi^0}_{\rm eff}\= \Big( |R_A^{\,\pi^0} + R_P^{\,\pi^0}|^2 +
       \frac{s}{4m^2}\, | R_P^{\,\pi^0}|^2\Big)^{1/2}\,.
\label{ff-eff}
\ee   
Lack of suitable polarization data prevents to disentangle both these
form factors anyway. The eqs.\ \req{hb-amp}, \req{cross-section} are
the $s \leftrightarrow t$ crossed versions of the handbag amplitudes
and the cross section for $\gamma p\to \pi^0 p$ derived in Ref.\
\ci{huang:04}. There are only some minor modifications for the form
factors occuring as a consequence of the specific frame of reference
used in the space- and time-like regions. These modifications which
have extensively been discussed in Ref.\ \ci{DKV3}, are: Instead of
vector and tensor generalized distribution amplitudes or form factors 
it is of advantage to apply the
Gordon decomposition to the vector piece in the time-like region
and trade the tensor form factor for the scalar one. It then turns out
that, in contrast  to the space-like region, the scalar form factor
decouples in the symmetric frame with $\zeta=1/2$ while the
pseudoscalar one contributes.\\ 
As is obvious from Eq.\ \req{handbag-amp} the pseudoscalar anihilation
form factor $R_P^{\pi^0}$ generates the $p\pb\to q\qb$ transitions 
where proton and antiproton have the same helicities while quark and
antiquark have opposite ones. This implies parton configurations of
the $p\pb$ system with non-zero orbital angular momentum. It is
therefore expected that, at large $s$, the pseudoscalar form factor is
suppressed as compared to the other form factors by at least
$1/\sqrt{s}$. If so it will not dominate over the other terms in 
Eqs.\ \req{cross-section}, \req{ff-eff} with increasing $s$.
    
\section{The annihilation form factors}
\begin{figure}
\begin{center}
\includegraphics[width=6.5cm,bb=107 365 463 723,clip=true]
{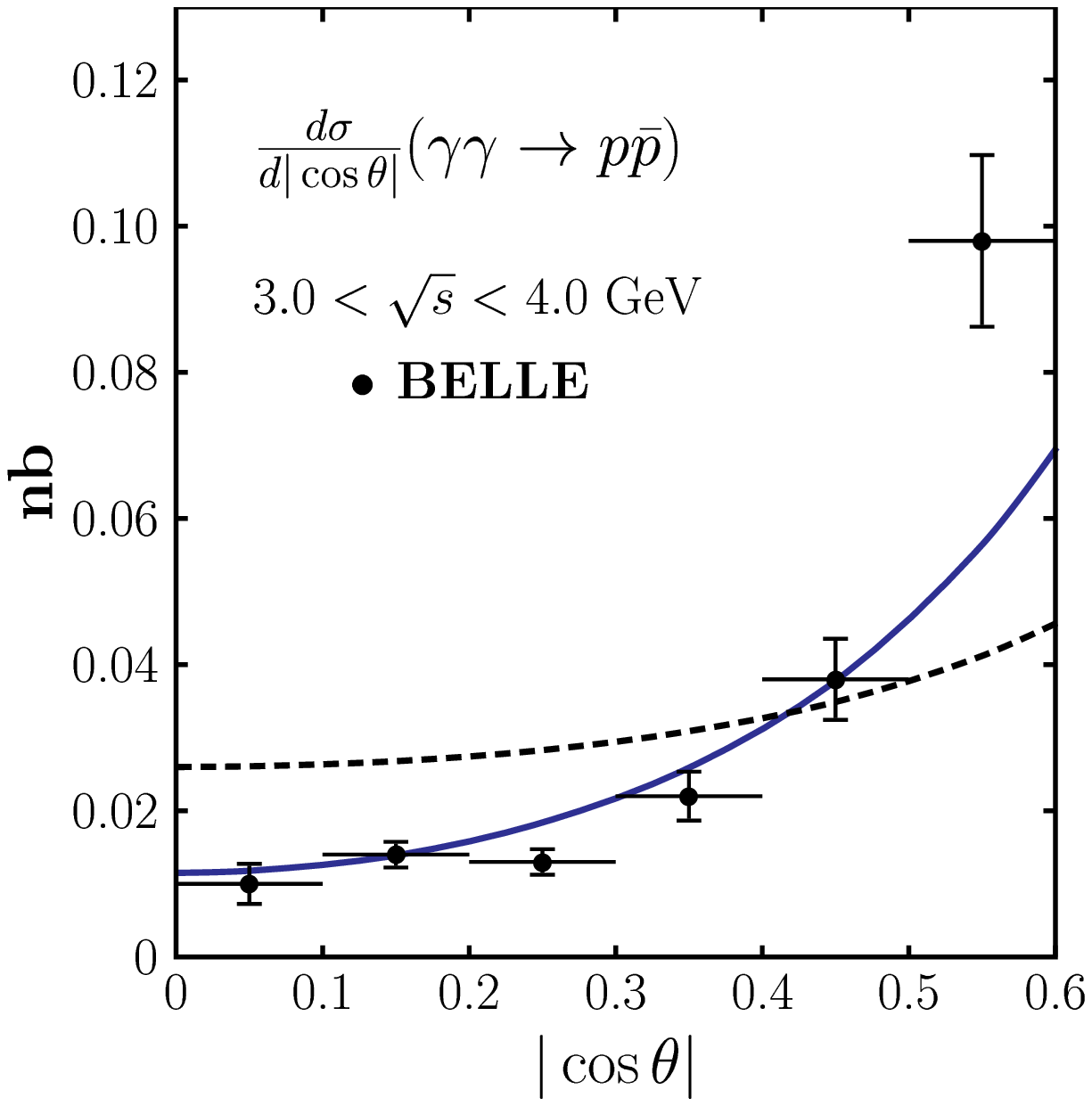}
\end{center}
\caption{The differential cross section for $\gamma\gamma\to p\pb$ 
versus $\cos{\theta}$ at $3.0 \leq \sqrt{s} \leq 4.0\,\gev$. 
Data taken from Ref.\ \ci{BELLE}. The solid line represents
our fit while the dashed one shows the angle dependence assuming
form factors, $R^{\,\gamma}_{\rm eff}=R^{\,\gamma}_V$, for comparison.}
\label{fig:ppbar}
\end{figure}

As is obvious from Eq.\ \req{hb-amp} an
analyis of the process $p\pb\to\gamma\,\pi^0$ at large $s,-t,-u$
requires information on the annihilation form factors \req{pi-ff}
which encode the physics of the soft $p\pb\to q\qb$ transition. 
Presently the form factors cannot be calculated from first principles
in QCD. Also there is no model calculation available for them currently.
Thus, we have to
determine the form factors phenomenologically.\\
Exploiting the universality property of the generalized \da s, we 
can use the form factors that have been determined from an analysis 
of two-photon annihilations into $p\pb$ pairs within the handbag 
approach \ci{DKV3}. We however do not use the results of the
numerical analysis presented in Ref.\ \ci{DKV3} since we have now at
our disposal the very accurate BELLE data on the differential and
integrated cross sections for $\gamma\gamma\to p \pb$ \ci{BELLE}. 
These data allow a determination of the vector and effective form factors 
from a
Rosenbluth type separation. We therefore redo the analysis of the
annihilation form factors. 
In the handbag approach \ci{DKV3} the $\gamma\gamma\to p \pb$ cross
section is given by an expression which  is analogous to the 
$|\overline{C}_2|$-term in Eq.\ \req{cross-section} with  form factors
defined analogously to Eqs.\ \req{pi-ff} and \req{ff-eff}:
\be
\frac{d\sigma}{dt}(\gamma\gamma\to p\pb)\= \frac{4\pi\alpha^2_{\rm
    elm}}{s^2}\, \frac1{\sin^2{\theta}}\, \left\{|R^{\,\gamma}_{\rm eff}|^2
    \,+\, |R_V^{\,\gamma}|^2\,\cos^2{\theta}\right\}\,.
\ee
A fit to the BELLE data on the differential cross section
at the highest measured energy ($3<\sqrt{s}<4\, \gev\,,$ implying an
average $s$ of $s_0=10.4\, \gev^2$) and on the integrated cross sections
for $s\geq 8\,\gev^2$ provides \footnote{
The cross section data in the region of $\eta_c$ formation is removed
from the fit. Possible signals from the $P$-wave charmonia are not
visible in the BELLE data.}$^,$ \footnote{
In this energy region the  BELLE data for the differential cross
section reveal a minimum at an scattering angle of $90^\circ$. For
lower energies the cross sections behave differently \ci{BELLE}.} 
\ba
s^2 R^{\,\gamma}_{\rm eff} \= (2.90\pm 0.31)\,\gev^4\, 
                            \left({s}/{s_0}\right)^{(-1.10\pm 0.15)}\,,\nn\\
|s^2 R^{\,\gamma}_{V}| \= (8.20\pm 0.77)\,\gev^4\, 
                            \left(s/s_0\right)^{(-1.10\pm 0.15)}\,, 
\label{gg-ff}
\ea
where the same energy dependence is assumed for both form factors.
The fit to the differential cross section is compared to experiment in
Fig.\ \ref{fig:ppbar}. The quality of the fit is, with the exception
of the data point at $\cos{\theta}=0.55$, very good. The failure with
this point may be regarded as an indication that the handbag approach
is here at its limits. Indeed for $s\simeq10.4\, \gev^2$ this
scattering angle corresponds to $t\simeq -2\,\gev^2$ which is not much
larger than $\Lambda^2$. For comparison we also show in Fig.\
\ref{fig:ppbar} the scattering angle dependence of the cross section
in the case that $R^\gamma_{\rm eff}$ equals $R^\gamma_{V}$ (this curve
is arbitrarily normalized). Evidently the form factors are different. 

The BELLE data exhibit clear evidence for violations of the
dimensional counting rules \ci{bro:73}, the integrated cross section
falls off faster than $s^{-5}$ resulting in an energy dependence of
the annihilation form factors as given in \req{gg-ff} instead of a
constant behaviour. As we mentioned above, due to Sudakov
suppressions the annihilation form factors are expected to decrease
faster than $1/s^2$ for very large $s$. 
Recent BELLE measurements \ci{belle-meson} 
on $\gamma\gamma \to \pi^+\pi^-$ and $K^+K^-$ as well as the 
JLab measurement \ci{bogdan} of Compton scattering also signal 
violations of the dimensional counting rules for hard scales of the
order of $10\,\gev^2$.\\ 
We note that the values for the form factors quoted in \req{gg-ff} are
somewhat different from the estimate given in Ref.\ \ci{DKV3}. This
estimate was based on the the CLEO and VENUS data \ci{CLEOp} which are
of markedly worse quality than the BELLE data \ci{BELLE} and were
only available at the rather low value $s=7.3\,\gev^2$.\\   
Since we have only protons at our disposal an exact flavor decomposition 
is not possible. We therefore follow Ref.\ \ci{DKV3} and simplify 
the expressions by taking a single real constant $\rho$ for the $d/u$ 
ratio of all proton form factors 
\be
F_i^{\,d} \= \rho\, F_i^{\,u}\,, \qquad i=V,A,P\,.
\label{ud-rel}
\ee
This ansatz parallels the behavior of fragmentation functions for
$d\to p$  and $u\to p$ transitions. In Ref.\ \ci{DKV3} the range of values 
\be
\rho\= 0.25 - 0.75
\ee
has been considered. We remark that simple quark counting arguments give  
$\rho=1/2$ \ci{kniehl:00}. Combining \req{ud-rel} with the flavor
decomposition of the form factors for $\gamma\gamma\to p\pb$
\be
R^{\,\gamma}_i\= e_u^2\,F^{\,u}_i + e_d^2\, F^{\,d}_i\,,
\ee
and neglecting non-valence quark contributions, one finds for the form
factors \req{pi-ff} of $\gamma\pi^0$ production
\be
   R_i^{\,\pi^0} \= \frac{1}{e_u \sqrt{2}}\,
   \frac{1-e_d/e_u\,\rho}{1+(e_d/e_u)^2\, \rho}\, R_i^{\,\gamma}\,, \qquad {\rm i= V,
     A, P}
\ee
Using the numerical values for the $R_i^\gamma$ quoted in \req{gg-ff}
and adding the errors of the form factors and $\rho$ quadratically, we
arrive at the estimate
\ba  
|s^2 R^{\,\pi^0}_{\rm eff}| \= (3.42\pm 0.4)\,\gev^4\, 
                            \left(s/s_0\right)^{-1.10\pm 0.15}\,,\nn\\
|s^2 R^{\,\pi^0}_{V}| \= (9.67\pm 1.1)\,\gev^4\, 
                            \left(s/s_0\right)^{-1.10\pm 0.15}\,. 
\label{gp-ff}
\ea
As we said above the effective form factor is 
likely dominated by the axial-vector
form factor since the pseudoscalar one involves parton orbital angular
momentum and is, therefore, expected to be suppressed as compared with
the other form factors at large $s$. An experimental separation of the 
axial vector and the pseudoscalar from factors requires polarization
experiments. The analysis of such observables is beyond the scope of
this work. We will only briefly comment on this issue below. For the
analysis of cross sections we only need the effective form factor.\\
Additional information on the vector form factor is provided by
the data on the magnetic proton form factor $G^{\,p}_M$ in the
time-like region for $s$ up to 14.4 \gev$^2$ \ci{E835:02}. This form
factor is related to the vector form factors \req{moments} by
\footnote{Note that only the form factors \req{moments} are universal
in contrast to the flavor combinations occuring in specific processes.}
\be
G^{\,p}_M(s) \= \sum_{a=u,d} e_a\, F_V^{\,a}(s) \,.
\ee 
The E835 data on the scaled magnetic form factor $s^2 |G^{\,p}_M|$ 
can also be represented by a power law
\be
|s^2 G^{\,p}_M|\= (2.46 \pm
0.16)\,\gev^4\,\left(s/s_0\right)^{(-0.75\pm 0.34)}\,.
\ee
An estimate of $R^{\,\pi^0}_V$ from $G^{\,p}_M$ along the same lines as
described above, leads to a smaller value than given in \req{gp-ff}
\ci{DKV3}. Given the approximations made in the derivation of the
handbag contribution \ci{DKV3}, there is however no real contradiction.

\section{Analysis of the Fermilab (E760) data}
\begin{figure}
\centerline{\includegraphics[width=7.5cm,bb=102 397 444 735,clip=true]
{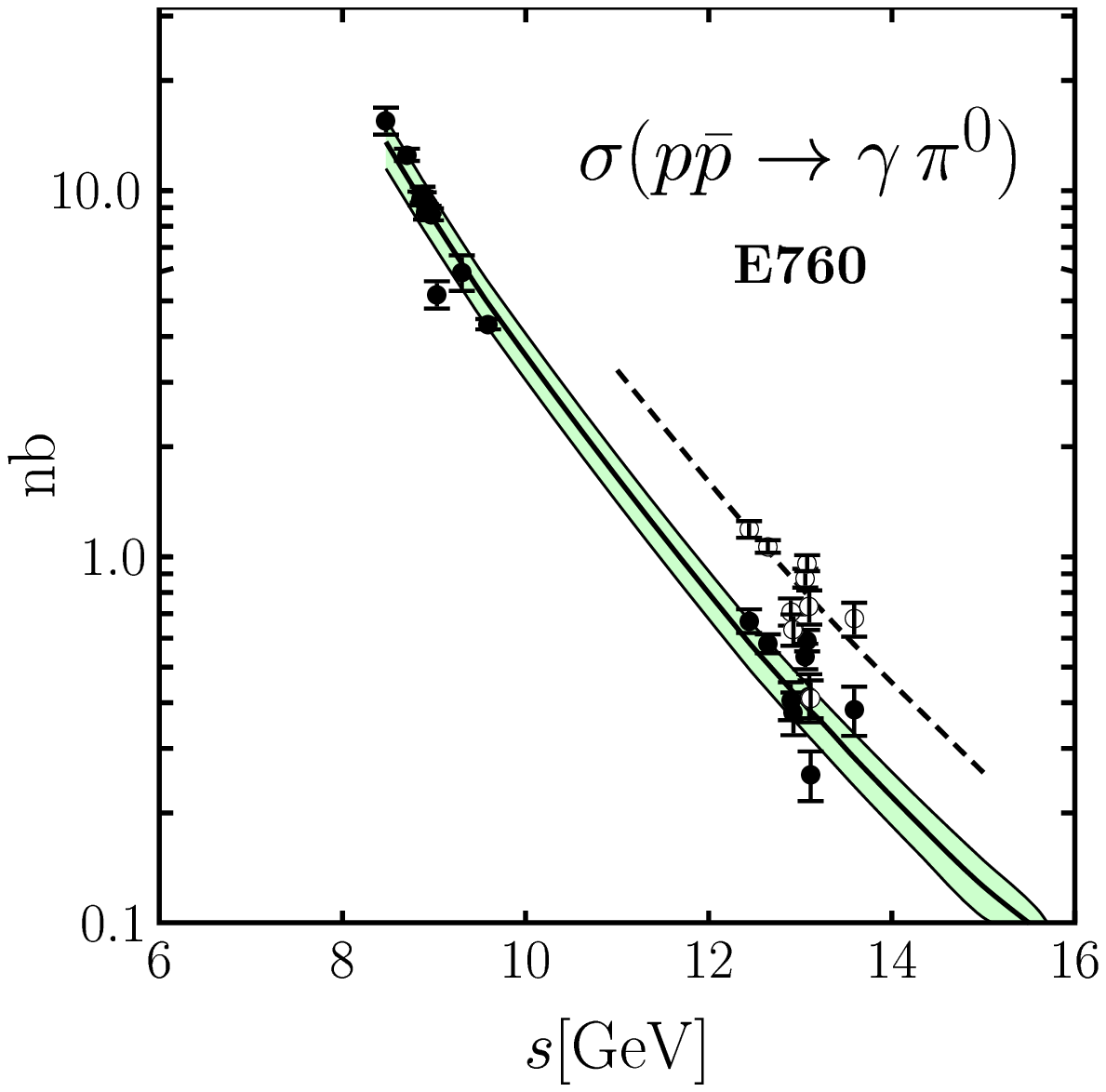}}
\caption{The integrated cross section for $p\pb\to\gamma\,\pi^0$
  versus $s$. Filled (open) circles: integrated over the range 
  $|\cos{\theta}|=0$ to $z_0$,  $z_0 = 0.5\; (0.6)$. 
  Data taken from \ci{E760:97}. The black line represents the
  prediction from the handbag approach with the error bands evaluated 
  from the uncertainties of the annihilation from factors. The dashed 
  line represents a power-law fit for $z_0=0.6$ with 
  $\sigma \sim (s/s_0)^{8.16\pm 0.12}$.}
\label{fig:cross-int}
\end{figure}

\begin{figure}
\begin{center}
\includegraphics[width=6.5cm,bb=100 360 465 694,clip=true]
{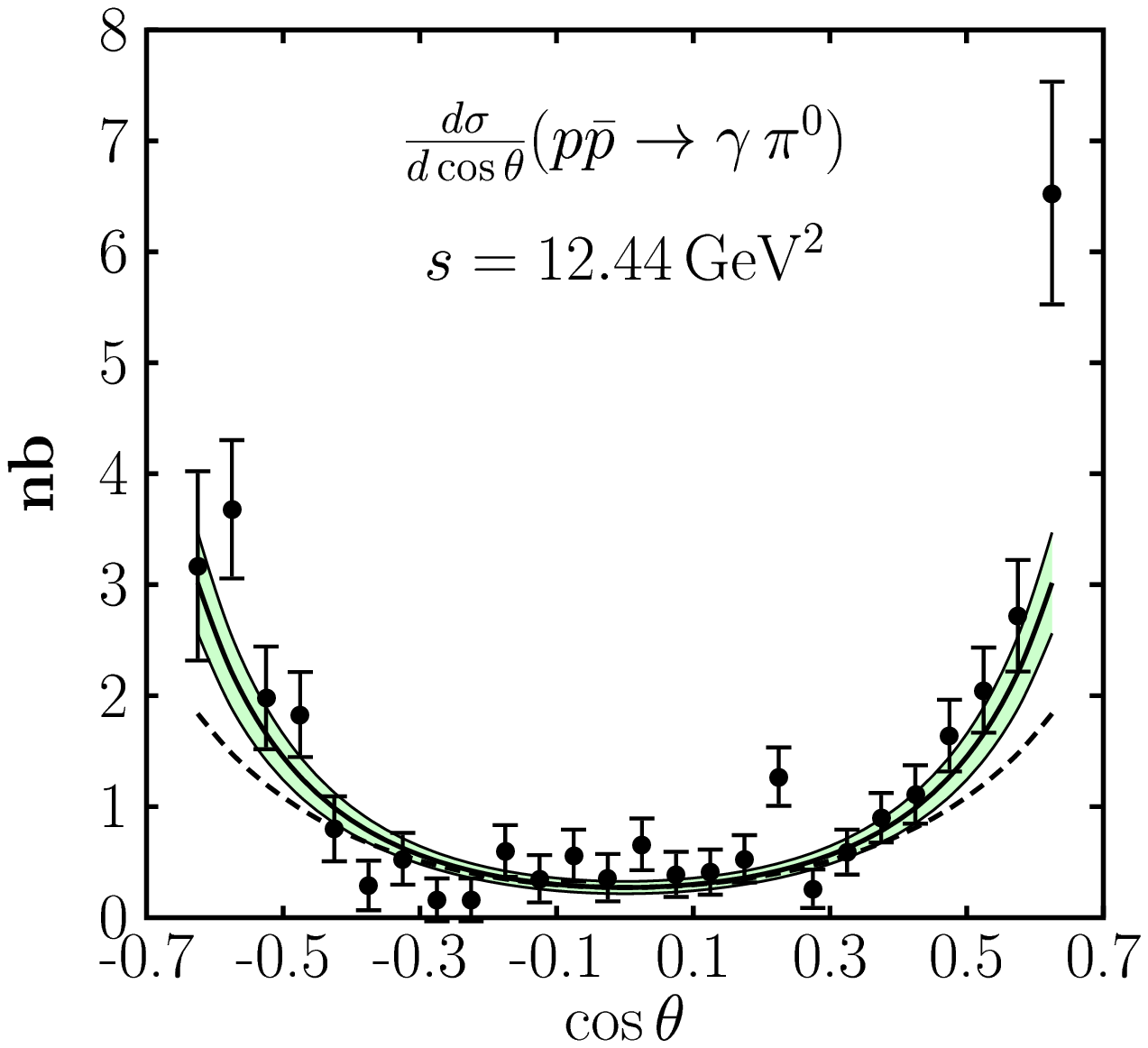}
\includegraphics[width=6.5cm,bb=109 385 476 718,clip=true]
{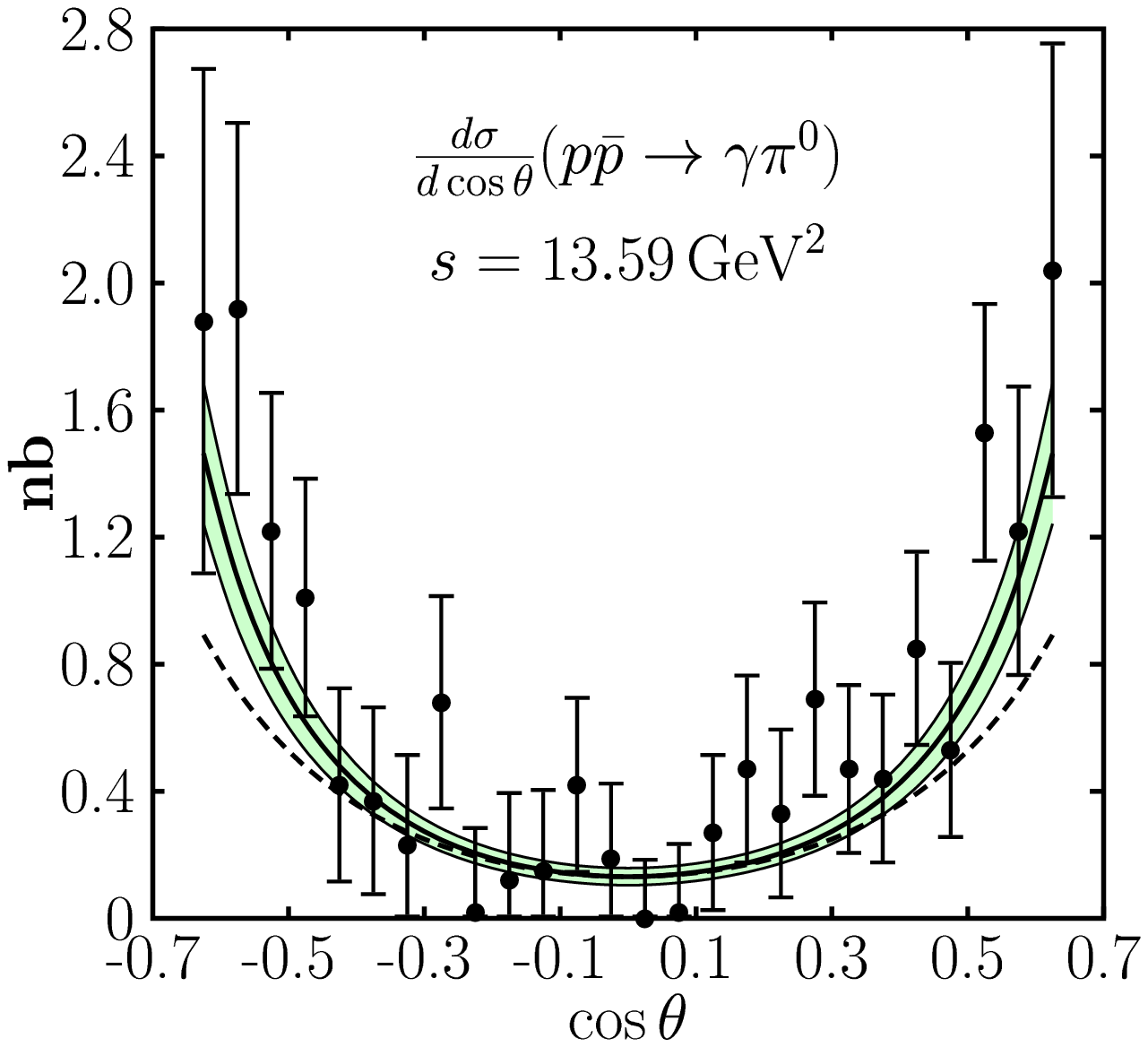}
\end{center}
\caption{The differential cross section for $p\pb\to\gamma\,\pi^0$ 
versus $\cos{\theta}$ at $s=12.44\,\gev^2$ (left) and $13.59\,\gev^2$ 
(right). Data taken from \ci{E760:97}. The solid lines with the error
bands represent the prediction from the handbag approach. For
comparison results are also shown where the cross section behaves 
$\propto 1/\sin^2{\theta}$ instead of $\propto 1/\sin^4{\theta}$ (dashed line).}
\label{fig:cross-diff}
\end{figure}

\begin{figure}
\begin{center}
\includegraphics[width=7.5cm,bb=118 398 458 734,clip=true]
{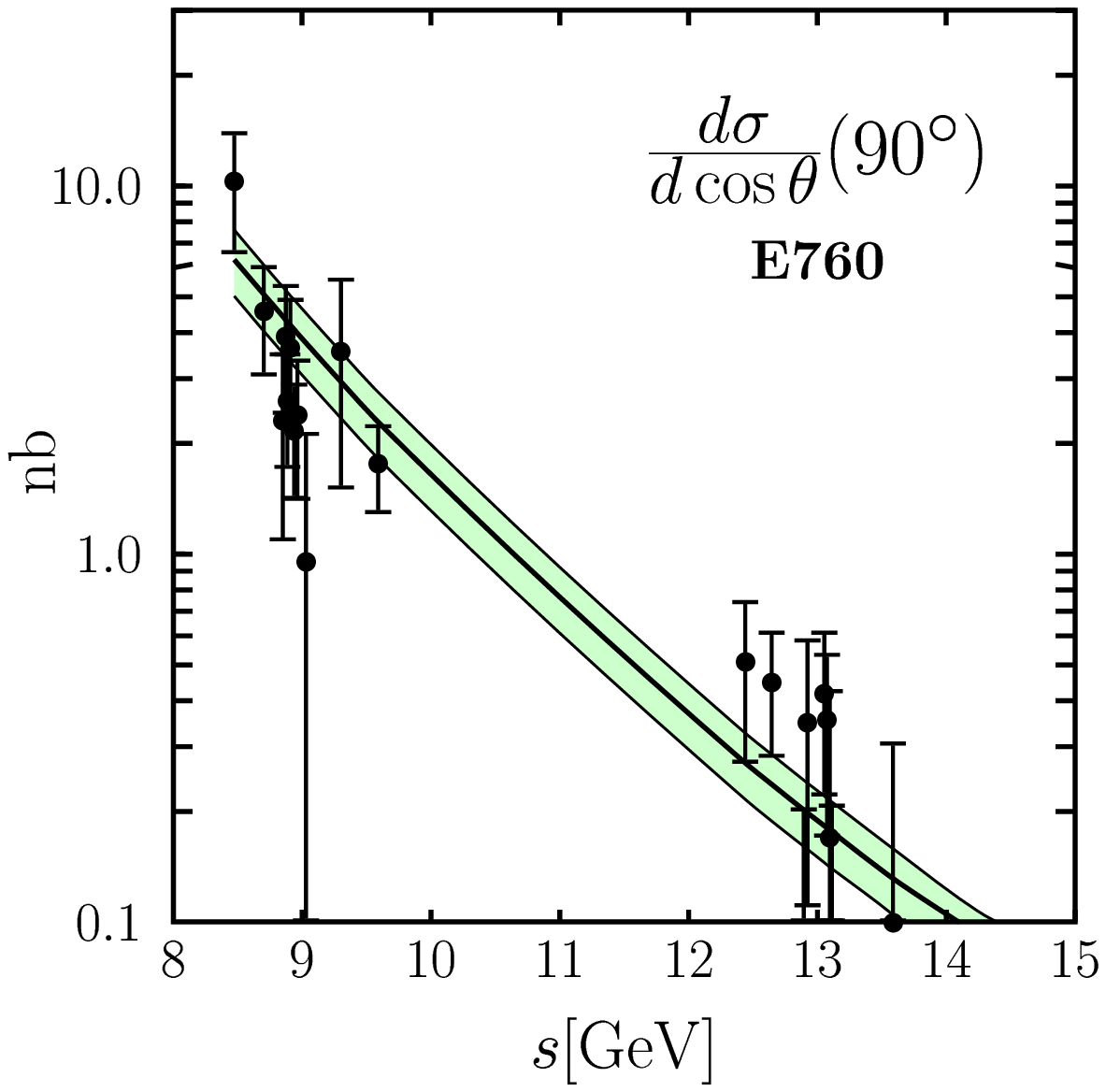}
\end{center}
\caption{The differential cross section for $p\pb\to\gamma\,\pi^0$ 
at $90^\circ$ versus $s$. Data, representing the average of the cross
sections in the two bins adjacent to $90^\circ$, are taken from
\ci{E760:97}. The solid line
with the error band represents the prediction from the handbag approach.}
\label{fig:cross-90}
\end{figure}

Let us turn now to the discussion of the CGLN invariant functions.      
They have definite behaviour under $t \leftrightarrow u$ crossing \ci{CGLN}:
\be
\overline{C}_2(\sh,\th) \= \phantom{-}\overline{C}_2(\sh,\uh)\,, \qquad
\overline{C}_3(\sh,\th) \= - \overline{C}_3(\sh,\uh)\,,
\label{crossing}
\ee
The invariant functions have dynamical
singularities. This can be seen from their leading-twist contributions
which are expected to dominate at large $s$.
In this kinematical region the
subprocess $q \qb \to \gamma \pi^0$ is dominated by 
one-gluon exchange to be calculated in collinear approximation to
lowest order of perturbative QCD. One obtains
\be
\overline{C}^{\,{\rm coll}}_2(\sh,\th) \= \frac{\bar{a}^{\,{\rm coll}}_2}{\th\uh}\,, \qquad
\overline{C}^{\,{\rm coll}}_3(\sh,\th) \= 0\,,
\label{eq-coll}
\ee
where 
\be
\bar{a}^{\,{\rm coll}}_2 \= 4\pi \alpha_s(\muR) f_\pi \frac{C_F}{N_c}
\langle{1/\tau}\rangle_\pi\,.
\label{a-coll}
\ee  
Here, $\muR$ is an appropriate renormalization scale, $f_\pi$ the pion
decay constant, $C_F=(N_c^2-1)/(2N_c)$ is the usual SU($N_c=3$) color 
factor and the last factor in \req{a-coll} is the $1/\tau$ moment of the 
pion distribution amplitude. To leading-twist accuracy also
$\overline{C}_1$ and $\overline{C}_4$ are zero. These invariant
functions control the situation where quark and antiquark have the
same helicity. They are only non-zero at twist-3 (or higher) level. 
As we mentioned in the introduction the leading-twist, lowest
order pQCD contribution to the subprocess falls short as compared
to experiment. It is therefore suggestive to assume that handbag 
factorization holds and that a more general mechanism than
leading-twist, unknown at present, is at work for the generation of
the meson and enhances the invariant function $\overline{C}_2$ sufficiently.\\ 
In view of this let us try now to find a suitable parametrization of
the invariant functions.
Their singularity structure will not be
altered by the inclusion of more complicated dynamical effects such as
higher-orders of perturbative QCD, transverse degrees of freedom
and/or higher twists or power corrections. For instance, by insertion 
of an infinite number of fermionic loops in the hard gluon propagator 
and by interpreting the ambiguities in the resummation of these loop 
effects - known as infrared renormalons - as a model of higher-twist 
contributions an enhancement of the leading-twist by a large factor 
may be obtained \ci{bel}.  Therefore, a general ansatz for
$\overline{C}_2$ reads
\be
\overline{C}_2(\sh,\th)\= \frac{\bar{a}_2}{\th\uh}\,f_2(\th,\uh)\,,
\label{ansatz}
\ee        
where $f_2$ is a symmetric function of $\th$ and $\uh$. It may
comprise powers of $\th\uh/\sh^2$, i.e.\
powers of $\sin{\theta}$, or
symmetric combinations of $\ln{\th}$ and $\ln{\uh}$. \\
From the leading-twist result \req{eq-coll} one may expect that the 
invariant function $\overline{C}_3$ plays only a minor role at $s$ of 
the order of $10\,\gev^2$ and may be neglected. This assertion is
further supported by an observation made for $\pi^{\pm}$
photoproduction \ci{huang:04}: the experimental ratio of these two 
processes is, within the handbag approach, only understood if 
$|{C}_2|\gg |{C}_3|$ provided quark helicity flip
can be neglected. Here, the $C_i$ are the $s\leftrightarrow t$ crossed
invariant functions $\overline{C}_i$. With regard to this observation we
will first discuss a scenario with a dominant $\overline{C}_2$
assuming for it the ansatz \req{ansatz} with the simplest choice $f_2
\equiv 1$. Taking the estimate \req{gp-ff} for the annihilation form
factors we are left with only one free parameter, namely the
normalization constant $\bar{a}_2$. In general $\bar{a}_2$ is a
complex number but the cross section only probes its modulus.
The energy and angle dependence of the
$p\pb\to\gamma\pi^0$ cross section is fixed by the handbag physics.\\
The differential and integrated cross section read for this scenario
($\theta$ is the angle between incoming proton and the outgoing
photon in the center of momentum frame.)
\ba
\frac{d\sigma}{d\cos{\theta}} &=& \frac{\ale}{4 s^6}
           \frac{|\bar{a}_2|^2}{\sin^4{\theta}}\, 
           \Big[ |s^2\,R^{\pi^0}_{\rm eff}|^2 
                     + \cos^2{\theta}\,|s^2\,R^{\pi^0}_V|^2 \Big]\,,\nn\\[0.5em]
\sigma  &=& \frac{\ale}{4}
         \frac{|\bar{a}_2|^2}{s^6} \left[\frac12\ln\frac{1+z_0}{1-z_0}
        \Big(|s^2\, R_{\rm eff}^{\pi^0}|^2-|s^2\, R_V^{\pi^0}|^2\Big)
                      \right.\nn\\
        &+& \left.  \frac{z_0}{1-z_0^2}\, 
        \Big(|s^2\, R_{\rm eff}^{\pi^0}|^2+|s^2\, R_V^{\pi^0}|^2\Big)\right]\,,
\ea
where $z_0=|\cos{\theta_0}|$ is the limit of integration.
With $z_0=0.5$ the integrated cross section is
\be
\sigma \= 864 {\rm nb}\,\gev^2\,
\frac{|\bar{a}_2|^2}{s^6}\,
                      \Big[|s^2\,R^{\pi^0}_{\rm eff}|^2 
                       + 0.096 |s^2\, R_V^{\pi^0}|^2\Big]\,.   
\ee
A fit to the E760 cross section data provides the value  
$(13.39\pm 0.10)\, \gev$ for the parameter $|\bar{a}_2|$. The fit is compared
to the data \ci{E760:97} in Fig.\ \ref{fig:cross-int}. With regard to
the errors of the annihilation from factors quoted in \req{gp-ff},
very good agreement of theory and experiment can be claimed.
The contribution from $J/\Psi$ formation is very small and can be
ignored. We also neglect a possible contribution from the
$\Psi^\prime$. Its decay into $\gamma\pi^0$ has not yet been observed.\\
For a few energies one may also integrate the E760 data up to $z_0=0.6$.
For comparison and in order to demonstrate the internal consistency of
our numerical analysis we fit a power law to these data. As can be
seen from Fig.\ \ref{fig:cross-int} the fit is in good agreement with
the data and yields the value of $8.16\pm 0.12$ for the power in
perfect agreement with the handbag approach and the form factors
\req{gp-ff} estimated from $\gamma\gamma\to p\pb$.  We note in passing
that a power of 6 is expected from the dimensional counting rules
our process.\\ 
The results from the handbag approach for the differential cross
sections are shown for two energies and for $|\cos{\theta}|\leq 0.6$ 
in Fig.\ \ref{fig:cross-diff}. The excellent agreement between the
theoretical results and the E760 data is obvious. Similarly good
results are obtained for all energies above $10\,\gev^2$ while for 
lower energies the cross section data exhibit marked fluctuations
which might be indicative of prominent low orbital angular momentum 
partial waves. This behaviour of the $p\pb\to \gamma\pi^0$ cross
section parallels that observed in $p\pb\to \gamma\gamma$
\ci{CLEOp,BELLE} and is the reason why we determined the vector
form factor only from BELLE data \ci{BELLE} for the cross section at
$s=10.4\,\gev^2$. 
In order to elucidate the structure of the handbag contribution
further we display the $p\pb\to \gamma\pi^0$ cross section at a
scattering angle of $90^\circ$ in Fig.\ \ref{fig:cross-90}. Also here
we observe fair agreement between the handbag approach and the data
although the errors are substantial.\\ 
In Fig.\ \ref{fig:cross-diff} we display for comparison also the
results of a calculation assuming $f_2=\sin{\theta}$ (~for simplicity
we retain the value of the parameter $\bar{a}_2$). In this case the angle 
dependence of the cross section coincides with that for $p\pb\to
\gamma\gamma$ but it seems to be too weak for the $\gamma\pi^0$
channel in contrast to the two-photon channel where it is in very good
agreement with the BELLE data for $|\cos{\theta}|\leq 0.5$. Thus, one
may conclude that, within errors, there is no need for $f_2 \neq
1$. In other words, an ansatz for the invariant function
$\overline{C}_2$ that resembles the singularity structure of the 
leading-twist result but with a strength adjusted to experiment, 
describes the data very well.\\
From Fig.\ \ref{fig:cross-90} it is also obvious that a scenario
$|\overline{C}_3|\gg |\overline{C}_2|$ is in conflict with
experiment. The zero of the $t \leftrightarrow u$ antisymmetric 
invariant function $\overline{C}_3$ at $90^\circ$ would lead to
a corresponding zero in the differential cross section which is not
seen with the exception of a few energies where a zero is possible
within errors \footnote{
Explicit fits of the $\overline{C}_3$ scenario to the 
differential cross section data, using for instance the ansatz
$\overline{C}_3= \bar{a}_3 \cos{\theta}/(tu)$, provide results which
are too small in the vicinity of $90^\circ$ as a consequence of this
zero. On the other hand, the fit to the integrated cross section is of
the same quality as in the $\overline{C}_2$ scenario (with
$|\bar{a}_3|=20.05\pm 0.15\, \gev$).} 
. It goes without saying that a small admixture of 
$\overline{C}_3$ to a dominant $\overline{C}_2$ cannot be ruled out.
There is yet another test of the internal consistency of the handbag
approach. Our process is related to photoproduction of pions by 
$s \leftrightarrow t$ crossing as we mentioned occasionally. The form
factors are functions of $t$ in the latter case and are known from a
recent ana\-ly\-sis of nucleon form factors exploiting the sum rules 
satisfied by generalized parton distributions \ci{DFJK4}. Using 
the invariant function related by $s\leftrightarrow t$ crossing
to   $\bar C_2$
\be
C_2(s,t) \= \frac{a_2}{su}\,,
\ee
where the Mandelstam variables are now those for photoproduction, one
finds fair agreement with the high-energy wide-angle SLAC data
\ci{anderson:76} if the parameter $a_2$ is adjusted to these
data ($a_2=20.3\,\gev  $). For very large $s$ the parameter 
$a_2$ should coincide with $|\bar{a}_2|=13.39\pm 0.10$~GeV
but at the actual $s$ values of order $10\,\gev^2$ this
discrepancy is not implausible. Typical differences between time- and 
space-like values of many quantities are of this size \ci{sterman}. 
It is, however,  important to realize that the photoprodcution data 
\ci{anderson:76} need confirmation. 

\section{Predictions for FAIR/GSI}
\begin{figure}
\begin{center}
\includegraphics[width=7.5cm,bb=90 350 500 800,clip=true]
{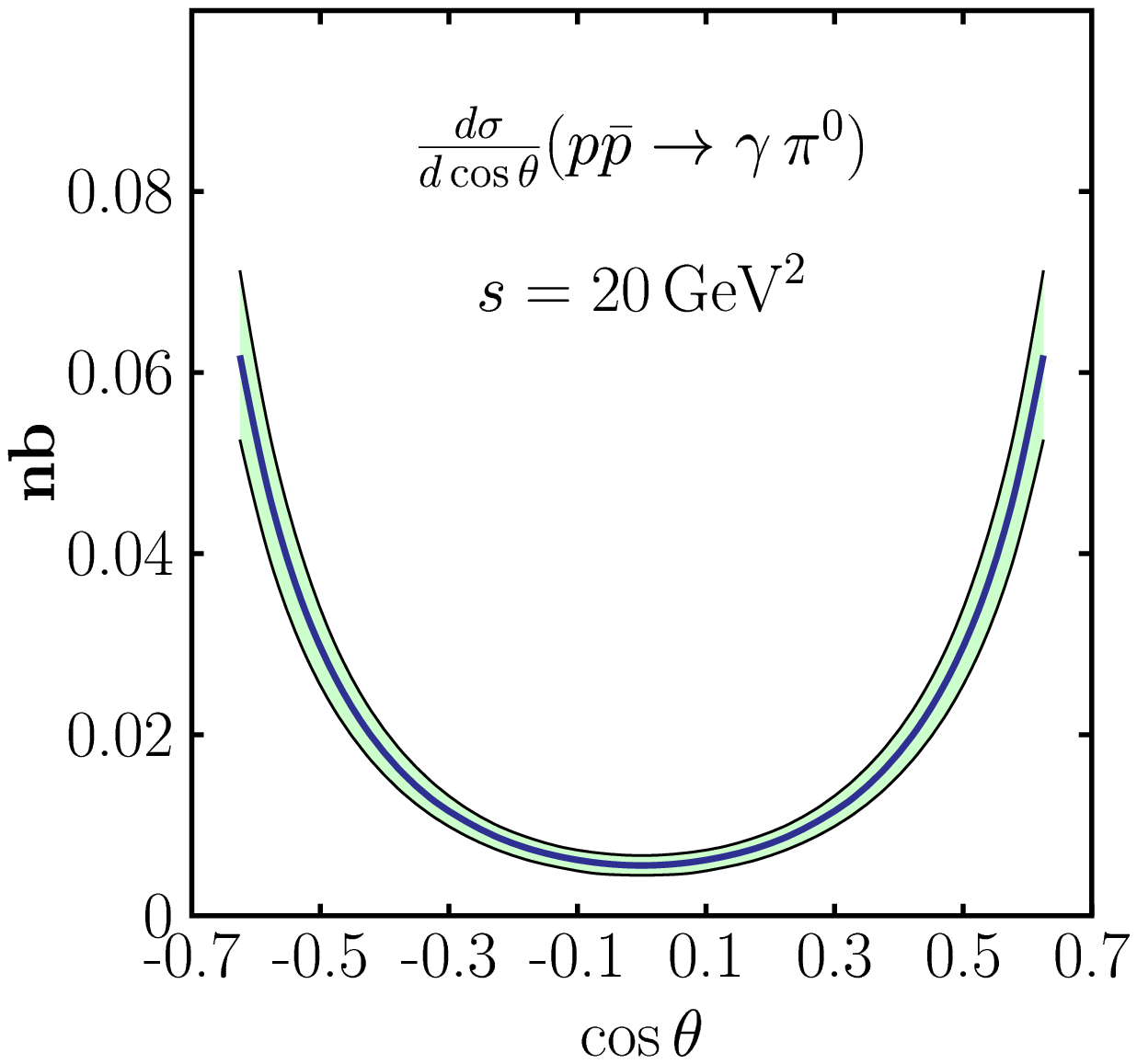}
\end{center}
\caption{Cross section prediction for the FAIR fixed-target energy 
$s=20$ GeV$^2$.} 
\label{fig:cross-pred}
\end{figure}

Obviously Figs.\ \ref{fig:cross-int} and \ref{fig:cross-90} provide
already predictions for the cross sections to be expected at FAIR. 
In Fig.\ \ref{fig:cross-pred} our prediction for the differential
cross section at $s=20$ GeV$^2$ is shown (the
fixed angle cross section scales as $s^{8.2\pm 0.3}$ in our approach).
As the discussed luminosities reach 
up to $2\cdot 10^4$ nb$^{-1}$/day  for unpolarized reactions   
and $10^3$ nb$^{-1}$/day for polarized ones, the data situation will 
improve drastically, ones FAIR is operational \cite{PAX}.\\  
The fact that the protons and antiprotons will be polarized 
adds very attractive additional possibilities. It allows e.g to
separate $R_A^{\pi^0}$ and $R_P^{\pi^0}$. In anlogy to the two-photon
channel \ci{DKV3} the helicity correlation $A_{LL}$ between proton and
antiproton is given by
\ba
A_{LL}&=& \frac{d\sigma(++)-d\sigma(+-)}{d\sigma(++)+d\sigma(+-)}\nn\\[0.2em]
      &=&-\, \frac{|R_A^{\pi^0}+R_P^{\pi^0}|^2 +\cos^2{\theta}|R_V^{\pi^0}|^2
                     -\frac{s}{4m^2}\,|R_P^{\pi^0}|^2}
            {|R_A^{\pi^0}+R_P^{\pi^0}|^2 +\cos^2{\theta}|R_V^{\pi^0}|^2
                   +\frac{s}{4m^2}\,|R_P^{\pi^0}|^2}\,,
\ea
where $d\sigma(\nu\nu')$ is the cross section for polarized
proton-antiproton annihilation.\\ 
Our analysis can be easily generalized to other pseudoscalar mesons, 
e.g. to the $\gamma\,\eta\, (\eta')$  channel.
The only complication arises from the additional
$s\sb$ and two-gluon Fock component the $\eta$ and $\eta'$ mesons
possess. For partons which are emitted nearly collinearly by the proton and
antiproton, each carrying a large fraction of its parent's momentum, the strange
quarks are likely strongly suppressed. Not much is known about the
two-gluon component. There is only some, not very precise information on its
leading-twist distribution amplitude from a next-to-leading order
analysis of the $\gamma \to \eta, \eta'$ transition form factors and
the inclusive $\Upsilon\to \eta'X$ decay \ci{passek}. These analyses
tell us that to leading-twist accuracy the contribution from the
two-gluon Fock component to $\eta'$ production may be sizeable while
it is strongly suppressed in the case of the $\eta$. Beyond the
leading-twist level the role of the two-gluon contribution is
unknown in the case of the $\eta'$. Since the $\eta$ meson is
dominantly a flavor octet state, however, its two-gluon component should be
suppressed in any case. In view of this one can treat the $\eta$ meson
in the quark flavor basis and, neglecting the strange component, regard
it as $\cos{\phi}\;\eta_q$ where $\phi$ is the usual mixing angle in
the quark-flavor basis and $\eta_q$ is the quark part of the 
$\eta$ wave function. The annihilation form factors for the
production of the isoscalar $\eta_q$ read
\be
R_{\,i}^{\,\eta_q} = \frac{1}{\sqrt{2}} \Big(\, e_u\, F_i^{\,u} +
                                           \,e_d\, F_i^{\,d} \Big)\,. 
\label{etaq-ff}
\ee
Their numerical values are somewhat smaller than those for the $\pi^0$
given in \req{gp-ff}. Up to this difference and a new value of the
parameter $\bar{a}_2$ the handbag approach then predicts the same
energy and angle dependence of the cross sections for the $\gamma\eta$
channel as for the $\gamma\pi^0$ one. Provided the two-gluon
component also plays only a minor role in the case of the $\eta'$ the
ratio of the $\gamma \eta$ and $\gamma\eta'$ cross sections is given
by
\be
\frac{d\sigma(p\pb\to\gamma\eta)}{d\sigma(p\pb\to\gamma\eta')} \=
\cot{\phi}\,.
\ee
In \ci{FKS1} the $\eta - \eta'$ mixing angle has been determined to be $39.3^\circ$. \\

Our results for the $\gamma\pi^0$ channel can also be straightforwardedly
generalized to the $\gamma V_L$ channel where $V_L$ is a
longitudinally polarized vector meson, if we  rely again on
valence quark dominance. The annihilation form factors for 
$\rho^0$ and $\omega$ production are the same as given in \req{gp-ff} 
and \req{etaq-ff}, respectively. In contrast $\phi$ meson production is
expected to be strongly suppressed because of the mismatch of the
proton and $\phi$ meson valence quarks. The cross section for the $\gamma
V_L$ channel reads
\be 
\frac{d\sigma^{V_L}}{d\cos{\theta}} = \frac{\ale}{4 s^6}\, 
           \frac{|\bar{a}^{V_L}_2|^2}{\sin^4{\theta}}\, 
           \Big[ |s^2\,R^{V_L}_V|^2 
                     + \cos^2{\theta}\,|s^2\,R^{V_L}_{\rm eff}|^2
                     \Big]\,.
\ee
The dependencies on the  vector and the effective form factors 
are reversed in this case 
as a consequence of parity invariance (see, for instance
Ref.\ \ci{huang:00}). We therefore expect a somewhat flatter angular
dependence, close to $1/\sin^4{\theta}$, for the vector meson channels
than for $\gamma \pi^0$. The isolation of longitudinally polarized
$\rho^0$ mesons might be possible with the planned PANDA detector
\ci{PAX}.
   
The generalization to transversally polarized vector mesons, $V_T$, is more
intricate. One has to consider the subprocesses $q\qb\to \gamma V_T$
with equal and opposite quark and antiquark helicities. In the first case
one has to introduce  new $p\pb$ distribution amplitudes and, hence, a
new set of associated annihilation from factors. These new
distribution amplitudes are time-like versions of the helicity-flip
GPDs introduced in Ref.\  \ci{diehl:01}. For opposite quark and
antiquark helicities, on the other hand, the formation of the
transversally polarized vector meson requires a different mechanism
than for longitudinally ones. It is beyond the scope of this work to
analyse the processes $p\pb\to\gamma V_T$.

\section{ Summary} 
We have analysed the reaction $p\bar p\rightarrow \gamma\pi^0$
for large scattering angles assuming handbag-dominance. We obtained
a rather satisfying description of the existing Fermilab data from
E760. Far more precise data can be expected from FAIR/GSI which
also provides the possibility to use polarization variables to separate 
the different contributions.  Such experiments would contribute to 
the determination of the generalized distribution amplitudes of the nucleon.
It would also allow to clarify the nature of the dominant reaction mechanisms 
as a function of cm energy $s$. We argue that a similar analysis can be 
performed without major problems 
for annihilation in a photon plus either another pseudoscalar meson or 
a longitudinal vector meson.

\section{Acknowledgement} 
We acknowledge helpful discussions with M.\ Diehl, M.\ D\"uren and C.\ Patrigiani
and thank the Institute for Nuclear Theory for hospitality, where this 
project was started during the programme 'GPDs and Exclusive Processes'.
This work was supported by BMBF and, in part, by the Integrated
Infrastructure Initiative ``Hadron Physics'' of the European Union,
contract No.\ 506078.


\begin{thebibliography}{99}
\bibitem{GPDs}
For recent reviews see:\\
K.~Goeke, M.~V.~Polyakov and M.~Vanderhaeghen,
  Prog.\ Part.\ Nucl.\ Phys.\  {\bf 47} (2001) 401
  [hep-ph/0106012];
M.~Diehl, Phys.\ Rept.\  {\bf 388} (2003) 41 [hep-ph/0307382].

\bibitem{DKV2} M.~Diehl, P.~Kroll and C.~Vogt,
Phys.~Lett.~{\bf B} 532, 99 (2002) [hep-ph/0112274].

\bibitem{DKV3} M.~Diehl, P.~Kroll and C.~Vogt,
Eur.\ Phys.\ J.\ C {\bf 26}, 567 (2003)
[hep-ph/0206288].

\bibitem{freund:02} A.~Freund, A.~V.~Radyushkin, A.~Sch\"afer and C.~Weiss,
Phys.\ Rev.\ Lett.\  {\bf 90}, 092001 (2003)
[hep-ph/0208061].

\bibitem{PAX} 
The most up-to-date informations can be found on the GSI web-page:
http://www.gsi.de/zukunftsprojekt/experimente/index.html\\
Of special interest for the actual context are the PAX and PANDA 
web-pages:\\
http://www.fz-juelich.de/ikp/pax/public\_files/tp\_PAX.pdf\\
http://www.gsi.de/zukunftsprojekt/experimente/hesr-panda/index.html

\bibitem{CLEOp} M.~Artuso {\it et al.}  [CLEO Collaboration],
Phys.\ Rev.\ D {\bf 50}, 5484 (1994);
H.~Hamasaki {\it et al.}  [VENUS Collaboration],
Phys.\ Lett.\ B {\bf 407}, 185 (1997).
G.~Abbiendi {\it et al.}  [OPAL Collaboration],
Eur.\ Phys.\ J.\ C {\bf 28}, 45 (2003)
[hep-ex/0209052].

\bibitem{BELLE} C.C.\ Kuo [BELLE collaboration], hep-ex/0503006.

\bibitem{huang:00} H. W. Huang and P. Kroll, 
Eur.\ Phys.\ J.\ {\bf C17}, 423 (2000), 
[hep-ph/0005318].

\bibitem{huang:04} H.~W.~Huang, R.~Jakob, P.~Kroll and K.~Passek-Kumericki,
Eur.\ Phys.\ J.\ C {\bf 33}, 91 (2004)
[hep-ph/0309071].

\bibitem{pion} R.~Jakob and P.~Kroll,
Phys.\ Lett.\ {\bf B315}, 463 (1993)
[Erratum-ibid.\ {\bf B319}, 545 (1993)]
[hep-ph/9306259];
S.~J.~Brodsky, C.~R.~Ji, A.~Pang and D.~G.~Robertson,
Phys.\ Rev.\ {\bf D57}, 245 (1998)
[hep-ph/9705221];
B.~Melic, B.~Nizic and K.~Passek,
Phys.\ Rev.\ {\bf D60}, 074004 (1999)
[hep-ph/9802204];
P.~Eden, P.~Hoyer and A.~Khodjamirian,
JHEP {\bf 0110} (2001) 040
[hep-ph/0110297].

\bibitem{CGLN} G.\ Chew, M.\ Goldberger, F.\ Low and Y.\ Nambu,
               Phys.\ Rev.\ {\bf 106}, 1345 (1957).

\bibitem{JLab1} L.~Y.~Zhu {\it et al.}  [JLab Hall A Coll.],
Phys.\ Rev.\ Lett.\  {\bf 91} (2003) 022003
[nucl-ex/0211009].

\bibitem{E760:97}T.~A.~Armstrong {\it et al.}  [Fermilab E760 Collaboration],
Phys.\ Rev.\ D {\bf 56}, 2509 (1997).

\bibitem{pire} B.~Pire and L.~Szymanowski,
[hep-ph/0504255].

\bibitem{brodsky} G.~P.~Lepage and S.~J.~Brodsky,
Phys.\ Rev.\ D {\bf 22}, 2157 (1980).

\bibitem{DFJK4} M.~Diehl, T.~Feldmann, R.~Jakob and P.~Kroll,
Eur.\ Phys.\ J.\ C {\bf 39}, 1 (2005)
[hep-ph/0408173].

\bibitem{farrar} G.~R.~Farrar, E.~Maina and F.~Neri,
Nucl.\ Phys.\ B {\bf 259}, 702 (1985)
  [Erratum-ibid.\ B {\bf 263}, 746 (1986)].

\bibitem{dixon} T.~C.~Brooks and L.~J.~Dixon,
Phys.\ Rev.\ D {\bf 62}, 114021 (2000)
[hep-ph/0004143].

\bibitem{DFJK1} M.~Diehl, T.~Feldmann, R.~Jakob and P.~Kroll,
Eur.\ Phys.\ J.\ {\bf C8}, 409 (1999)  [hep-ph/9811253].

\bibitem{diehl:01} M.~Diehl,
Eur.\ Phys.\ J.\ {\bf C19}, 485 (2001)
[hep-ph/0101335].

\bibitem{DFHK} M.~Diehl, T.~Feldmann, H.W.~Huang and P.~Kroll,
Phys.\ Rev.\ D{\bf 67}, 037502 (2003) [hep-ph/0212138].

\bibitem{bro:73} S.~J.~Brodsky and G.~R.~Farrar,
Phys.\ Rev.\ Lett.\  {\bf 31}, 1153 (1973);\\
%
V.~A.~Matveev, R.~M.~Muradian and A.~N.~Tavkhelidze,
Lett.\ Nuovo Cim.\  {\bf 7}, 719 (1973).

\bibitem{belle-meson} H.~Nakazawa {\it et al.}  [BELLE Collaboration],
Phys.\ Lett.\ B {\bf 615}, 39 (2005)
[hep-ex/0412058].

\bibitem{bogdan} E99-114 JLab collaboration, spokespersons
C.\ Hyde-Wright, A.\ Nathan and B.\ Wojtsekhowski9999

\bibitem{kniehl:00}B.~A.~Kniehl, G.~Kramer and B.~P\"otter,
Nucl.\ Phys.\ B {\bf 582}, 514 (2000)
[hep-ph/0010289];
F.~E.~Close and Q.~Zhao,
Phys.\ Lett.\ B {\bf 553}, 211 (2003)
[hep-ph/0210277].

\bibitem{E835:02} M.~Ambrogiani {\it et al.}  [E835 Collaboration],
Phys.\ Rev.\ D {\bf 60}, 032002 (1999).
                 
\bibitem{bel} A.~V.~Belitsky,
AIP Conf.\ Proc.\  {\bf 698}, 607 (2004)
[hep-ph/0307256].



\bibitem{anderson:76} R.~L.~Anderson {\it et al.},
Phys.\ Rev.\ D {\bf 14}, 679 (1976).

\bibitem{sterman} L.~Magnea and G.~Sterman,
Phys.\ Rev.\ D {\bf 42} (1990) 4222;
A.~P.~Bakulev, A.~V.~Radyushkin and N.~G.~Stefanis,
Phys.\ Rev.\ D {\bf 62}, 113001 (2000)
[hep-ph/0005085].


\bibitem{passek}P.~Kroll and K.~Passek-Kumericki,
Phys.\ Rev.\ D {\bf 67}, 054017 (2003)
[hep-ph/0210045];
A.~Ali and A.~Y.~Parkhomenko,
Eur.\ Phys.\ J.\ C {\bf 30}, 183 (2003)
[hep-ph/0304278].

\bibitem{FKS1}T.~Feldmann, P.~Kroll and B.~Stech,
Phys.\ Rev.\ D {\bf 58}, 114006 (1998)
[hep-ph/9802409].
\end{thebibliography}
\end{document}